\documentclass[acmsmall]{acmart}

\acmJournal{TOSEM}
\acmVolume{1}
\acmNumber{1}
\acmArticle{1}
\acmYear{2025}
\acmMonth{10}

\setcopyright{acmcopyright}
\acmDOI{XXXXXXX.XXXXXXX}

\usepackage{graphicx}
\usepackage{subcaption}
\usepackage{booktabs}
\usepackage{amsmath,amsthm}
\usepackage{multirow,tabularx,array,float}
\usepackage[table,xcdraw]{xcolor}
\usepackage{tcolorbox}
\usepackage{svg}
\usepackage{framed}
\usepackage{fancyvrb}
\usepackage{hyperref}
\usepackage[capitalize,noabbrev]{cleveref}
\usepackage{soul}
\usepackage{tikz}
\usepackage{microtype}
\usepackage[textsize=tiny]{todonotes}

\newtcolorbox{summarybox}{
  colback=gray!10,
  colframe=black,
  boxrule=0.5pt,
}

\title{Test Case Generation from Bug Reports via Large Language Models: A Cognitive Layered Evaluation Framework}

\author{Irtaza Sajid Qureshi}
\email{irtaza11@yorku.ca}
\orcid{0000-0002-XXXX-XXXX}
\affiliation{%
  \institution{York University}
  \department{Lassonde School of Engineering}
  \city{Toronto}
  \country{Canada}
}

\author{Zhen Ming (Jack) Jiang}
\email{zmjiang@yorku.ca}
\orcid{0000-0001-XXXX-XXXX}
\affiliation{%
  \institution{York University}
  \department{Lassonde School of Engineering}
  \city{Toronto}
  \country{Canada}
}

\begin{abstract}
Large Language Models (LLMs) are increasingly applied to automated software testing, yet their ability to generalize beyond memorized patterns and reason about natural language bug reports remains unclear. We present a systematic evaluation of LLM reasoning in test case generation, structured around the cognitive layers of Bloom’s taxonomy: \textit{Remember}, \textit{Understand}, \textit{Apply}, \textit{Analyze}, \textit{Evaluate}, and \textit{Create}, which progressively assess higher levels of cognitive and reasoning capabilities. Building on the LIBRO framework, we evaluate StarCoder and GPT-4o on Defects4J, GHRB, and mutated variants that introduce linguistic and semantic challenges. Our findings show that both models largely reproduce prior results with minor deviations (\textit{Remember}), exhibit partial robustness to linguistic rephrasings and translations while uncovering unique reproducible bugs (\textit{Understand}), but suffer severe performance drops exceeding 60\% under identifier mutations (\textit{Apply}). Conversely, providing near-identical few-shot examples in an open-book setting improves success rates by up to three times, and component-level analysis reveals that structured technical elements, such as test code and method names, are far more impactful than narrative descriptions for successful test generation (\textit{Analyze}). These insights illuminate the cognitive processes underlying LLM-generated tests, suggest concrete directions for improving performance, and establish a robust and realistic evaluation paradigm for this task.
\end{abstract}

\begin{document}

\maketitle

\section{Introduction}
Large Language Models are rapidly evolving, with expanding use cases across software engineering. Their ability to generate code for well-described tasks makes them instrumental in automated development workflows. Among various applications of LLMs in this domain, one major area is software testing. LLMs can leverage natural language descriptions in bug reports to generate test cases aimed at reproducing the reported issues. The performance of LLMs on such tasks is often assessed using standardized benchmarks designed to evaluate their capabilities across different scenarios. A widely recognized benchmark for test case generation is Defects4J~\cite{Just_2014_D4J}, which has long served as a baseline for evaluating traditional techniques such as EvoCrash~\cite{Soltani_2020_EvoCrash} and Copy\&Paste~\cite{Bettenburg_2008_ESIfBR}.

However, the suitability of these benchmarks for evaluating LLM-based methods raises new concerns. Real-world bugs from open-source repositories, which form the basis of benchmarks like Defects4J, may have been included in the training corpora of modern LLMs. This overlap introduces the risk of data contamination, where models may reproduce memorized examples instead of demonstrating true generalization. Such contamination can inflate performance metrics and compromise the reliability of evaluation outcomes.

To address this issue, we propose a structured evaluation framework that systematically examines LLMs across progressively challenging scenarios. Our approach draws inspiration from Bloom's Taxonomy~\cite{Bloom_1956_Taxonomy}, a well-established model for categorizing cognitive skills that was originally designed for educational settings. We incorporate its layers: \textit{Remember}, \textit{Understand}, \textit{Apply}, \textit{Analyze}, \textit{Evaluate}, and \textit{Create}, to assess LLM-powered test generation along dimensions such as retention, generalization, contextual reasoning, ranking ability, and the capacity to recognize ambiguity and request clarifications.

Rather than proposing entirely new benchmarks every few years, which requires significant manual effort and data curation, our framework adopts dynamic benchmarking by extending existing benchmarks through systematic linguistic and contextual variations. This enables evaluation under conditions that simulate unseen scenarios while maintaining compatibility with prior work. Each layer in the framework targets a specific capability, from reproducing test cases under identical inputs to handling reworded bug reports, masked identifiers, and open-book prompts. In doing so, we evaluate whether LLMs can move beyond memorization and demonstrate robust, task-relevant reasoning.

We apply this framework to two modern LLMs, StarCoder and GPT-4o, using Defects4J and the GitHub Recent Bugs (GHRB) benchmark introduced by the LIBRO study~\cite{Kang_2023_LLMaFST}. Our case study spans more than 800 bugs and their corresponding variants, measuring both bug reproduction success and model consistency across layers. The results highlight where current models succeed, where they fall short, and how evaluation methodologies can better reflect real-world usage.

\textbf{Contributions.} This paper makes the following contributions:

\begin{itemize}
    \item We propose a structured evaluation framework for LLM-based test generation, inspired by Bloom’s Taxonomy, to assess capabilities such as retention, generalization, contextual reasoning, and clarification-seeking.
    
    \item We design controlled mutation strategies that systematically vary bug reports and code contexts, enabling robust evaluation across linguistic and semantic dimensions.

    \item We conduct a large-scale empirical study using Defects4J and GHRB to evaluate StarCoder and GPT-4o, offering detailed insights into their behavioral patterns across layered tasks.
\end{itemize}

\textbf{Paper Organization.} Section~\ref{sec:background} presents background and related work. Section~\ref{sec:methodology} describes our evaluation framework and methodology. Section~\ref{sec:casestudy} presents the results of our case study. Section~\ref{sec:discussion} discusses key findings and implications. Section~\ref{sec:conclusion} concludes the paper.

\section{Background and Related Work}
\label{sec:background}
In this section, we review the foundational work and recent advances related to our study, focusing on three key areas: test case reproduction, LLM-based test generation, and benchmark construction and evaluation in software engineering.

\subsection*{1) Test Case Reproduction}

Test case reproduction involves the automatic generation of tests that expose known software bugs. Traditional approaches, such as search-based software testing demonstrated by EvoSuite~\cite{Fraser_2011_EvoSuite}, employ genetic algorithms to evolve Java JUnit test suites toward satisfying coverage criteria. These methods optimize for coverage and fault detection by iteratively refining candidate test suites. While effective, such techniques typically require detailed access to program semantics and do not directly utilize natural language bug reports.

More recently, learning-based methods have begun to incorporate bug report information to guide test generation. These approaches improve adaptability to human-written bug reports, allowing automated systems to better interpret natural language. This accelerates test generation and supports quicker fault detection for faster bug resolution.

\subsection*{2) LLM-Based Test Case Generation}

LLM-based test case generation techniques leverage the powerful language understanding and code synthesis capabilities of LLMs to translate natural language bug reports into executable tests. LIBRO~\cite{Kang_2023_LLMaFST} employs few-shot prompting by providing the model with pairs of example bug reports and their corresponding tests. Then prompts the LLM to generate a test for a new bug report, which is validated by execution on both buggy and fixed program versions to ensure correct failure reproduction. Post-processing filters and ranks the generated tests to improve reliability.

Beyond LIBRO, other approaches explore different strategies to enhance test generation quality. These approaches apply few-shot prompting in multi-step pipelines comprising context localization, scaffolding, model planning, and feedback loops to improve the efficiency and accuracy of test case generation ~\cite{Ahmed_2024_TTDBench} ~\cite{Ahmed_2025_Otter} ~\cite{Nashid_2025_Issue2Test}. In addition, tools like ~\cite{Sapozhnikov_2024_TestSpark} integrate LLM-based test generation with traditional techniques (EvoSuite) within an IntelliJ plugin, enabling interactive test generation and refinement.

\subsection*{3) Software Engineering Benchmark and Evaluation}

Effective evaluation of test generation frameworks requires robust, realistic benchmarks that accurately reflect model capabilities and limitations. Defects4J~\cite{Just_2014_D4J} remains one of the most widely used benchmarks for Java-based test case reproduction, providing real bugs and fixes drawn from open-source projects. While it has enabled reproducible and comparative evaluation, its static and dated nature introduces significant limitations. Many of the repositories included in Defects4J may have been seen by models during pretraining, raising concerns about data contamination. This undermines the reliability of reported performance metrics and obscures the model’s actual reasoning abilities.

Benchmarks like SWT-Bench~\cite{Mundler_2025_SWTBench} and TDD-Bench~\cite{Ahmed_2024_TTDBench} offer more contemporary evaluation datasets by focusing on Python projects with real-world bug-fix tasks. While they reduce the likelihood of contamination in the short term, their static nature means they will eventually become part of future training data. This limitation highlights the ongoing need for dynamic and adaptive evaluation methods.

Beyond static benchmarks, \textit{dynamic benchmarking} has emerged as a promising direction for assessing LLM robustness and reasoning. Rather than relying on a fixed set of tasks, dynamic benchmarks generate novel, out-of-distribution, or mutated test scenarios adaptively. This approach addresses two key challenges: (1) mitigating contamination by ensuring evaluation data remains unseen and unpredictable, and (2) enabling fine-grained assessments that probe reasoning, problem-solving, and adaptability under evolving conditions. Similar methodologies have been explored in related domains, for example, frameworks like DyVal~\cite{Zhu_2024_DyVal} and DARG~\cite{Zhang_2024_DARG} construct and perturb structured reasoning graphs to create diverse, challenging test samples. While these ideas have gained traction in evaluating general LLM capabilities and some automated software engineering tasks, their systematic application to test case generation remains largely unexplored. Advancing beyond static datasets toward dynamic, contamination-resistant benchmarks will be key to ensuring fair, realistic, and forward-looking assessments.

\section{Methodology}
\label{sec:methodology}
Our methodology is designed as an experimental framework to assess the capabilities of LLM beyond the generic benchmark evaluation, which could be inflated due to training data contamination. This study systematically evaluates the generalization and memorization capabilities of LLMs in generating test cases from bug reports. Our framework is inspired by Bloom's Taxonomy, which is traditionally used in education research to quantify levels of learning and skill mastery, ranging from basic recall to complex problem-solving and knowledge transfer. We systematically divide the evaluation into six layers: Remember, Understand, Apply, Analyze, Evaluate, and Create. These layers are arranged in an increasingly complex manner, starting from Remember and progressing up to Create. Each layer is targeted to analyze different capabilities of the LLM by altering certain parts of the bug reproducing framework. To show the effectiveness of our approach, we demonstrate our study using Defects4J version 2.1.0. The rest of this section breaks down each layer in detail.

\subsection{Remember}
    \textbf{Objective:}
    The goal of evaluation at this layer is to understand whether the solutions under evaluation (SUEs) reproduce the results under the same experimental configurations.
    \\
    \textbf{Approach:}
    Evaluating the consistency of LLMs across model versions is crucial for understanding their reliability in automated test generation. Before assessing generalization to new scenarios, it is important to determine how well LLMs retain prior knowledge and reproduce results under identical conditions. In our evaluation framework, \textit{remembering} refers to the ability to recall and reproduce. We apply this concept to LLM evaluation by investigating whether a model consistently generates the same test cases when re-evaluated with the same framework. To do so, we replicate the LIBRO experimental setup with StarCoder-15.5B and examine whether results align with prior findings. Additionally, we extend this replication to GPT-4o, evaluating whether test case generation capabilities are preserved across model updates.
    \\
    \textbf{Evaluation Scheme:} 
    Similar to LIBRO, we track the percentage of bugs with \textit{bug-reproducing test cases (BRT)}. A BRT is a test case that compiles, fails on the buggy version, and passes on the fixed version. This serves as the primary indicator of successful bug reproduction. This evaluation helps determine whether the SUE can recall and regenerate valid test cases under identical conditions.

\subsection{Understand}
    \textbf{Objective:}
    The goal of evaluation at this layer is to check whether the SUEs can correctly interpret bug reports when the code context remains the same.
    \\
    \textbf{Approach:} 
    \textit{Understanding} a bug report requires an LLM to extract relevant information irrespective of phrasing or language. In this layer, we introduced linguistic variations through rephrasing and translating bug reports. We applied these variations to verify if these SUEs have basic comprehension skills to extract the key information from the bug reports, regardless of the phrasing. These two types of rephrasing were applied: 
    \
    \begin{itemize}
        \item \textit{Rephrasing:} Rephrasing was achieved by prompting GPT-3.5-turbo to linguistically vary the bug report but keep the original meaning intact. Each rephrased bug report was verified for semantic integrity with the original bug report by prompting GPT-4o. Rephrasing was performed again until the verification passed.
        
        \item \textit{Translating:} Similar to rephrasing, we extended our approach to include translations into Romanized Urdu to evaluate multilingual capabilities. We used Romanized Urdu, rather than a non-Latin script, to ensure compatibility with technical tools and enhance accessibility for testing frameworks. The translations were conducted using prompts to GPT-4 by preserving the meaning of the original bug report. Code snippets, API names, and stack traces were maintained in the original language. After manual analysis of a small sample of 5 bug reports from each project to ensure quality, GPT-4o was used for verification of the remaining bug reports. Figure \ref{linguistic_variation} illustrates how linguistic variations are applied to enhance the robustness of the evaluation.
    \end{itemize}
    \
    \textbf{Evaluation Scheme:} 
    As in the Remember layer, we track the percentage of bugs with BRTs. For each bug in a rephrasing setup, the SUE either solves it or does not. We result in a binary vector for each evaluation round that reflects success across all instances under a given setup. We used McNemar’s test to determine whether the performance of the SUEs differs significantly between the Remember and Understand layers, and across different rephrasing setups. In addition, we also compute the \textit{Baseline Consistency Rate}, which reflects the percentage of bugs successfully reproduced both in the baseline and the modified setting. This helps assess whether the SUE can maintain performance when bug reports or code are altered.
    
\subsection{Apply}
    \textbf{Objective:}
    The goal of evaluation at this layer is to examine the ability of SUEs to apply learned knowledge to different unseen scenarios, specifically when method names are altered through identifier mutations.
    \\
    \textbf{Approach:}
    Applying learned knowledge in test case generation requires that LLMs reason beyond memorized patterns and adapt to variations in code context. Many LLMs appear to be aligned with the identifier naming conventions of the repositories on which they were trained. When method names are altered, the models often experience performance declines. This reduction suggests that the models’ success is partially driven by their familiarity with specific identifiers, rather than by a deeper understanding of code semantics. To prevent LLMs from leveraging potentially memorized identifiers in their training data, we generated new method names for the bug-relevant methods using these strategies:
    \begin{itemize}
        \item \textit{Meaningful Masking:} We prompted GPT-3.5-turbo with relevant context from the method implementation to generate alternative method names. These names differ from those seen during training, requiring the SUEs to rely on contextual understanding rather than memorized identifiers when reasoning about the bug.
        
        \item \textit{Hash based Masking:} Instead of meaningful names, which can provide contextual information of the purpose of the function, we replace them with hash-based names by calculating the SHA-256 hashes for the original method names. This strips away all semantic meaning, forcing the SUEs to rely solely on the code's structure and behavior to understand bugs. This tests whether they can understand the bug purely through the logic of the implementation.
    \end{itemize}
    \
    Experiments in this layer were conducted using four configurations: hash masking, meaningful masking, rephrasing combined with meaningful masking, and Romanized Urdu combined with meaningful masking. These variations allow us to test whether SUEs can still apply learned knowledge when familiar identifiers and phrasing are modified.
    \\
    Given the buggy-fixed pair for the project under test, we use a \emph{diff} tool to identify patched methods. The original method names are anonymized using hash masking or meaningful mask. The generated method names are appended to the prefix “func\_” to ensure the new name is unique and compatible with Java identifier rules (i.e., not starting with a number). These new method names replace the original ones in the code base using the \textit{javalang} parser to identify method declarations and invocations.
    \\
    \textbf{Evaluation Scheme:}
    Similar to the Understand layer, we track the percentage of bugs with BRTs. We also report the \textit{Baseline Consistency Rate} to assess performance retention relative to the original setup. In addition, we also employed McNemar’s test to check whether the evaluation results differ statistically from the previous layers.

\subsection{Analyze}
    \textbf{Objective:}
    The goal of evaluation at this layer is to investigate how SUEs extract meaningful information from few-shot examples and bug report components, assessing their capacity to adapt retrieved examples for generating successful test cases.
    \\
    \textbf{Approach:} 
    Effective test generation requires more than recognizing patterns. It also demands the ability to analyze and extract relevant information from examples and bug reports. In this layer, \textit{analysis} refers to how well SUEs can learn from provided context and apply that knowledge to bugs under consideration. We evaluate this by comparing two settings: a closed-book setup using static few-shot examples curated by LIBRO researchers, and an open-book setup where examples are dynamically retrieved based on bug report similarity using BM25. The open-book setting introduces near-identical examples drawn from the same dataset, testing whether SUEs can recognize and adapt contextually similar cases to improve test generation.
    \
    \begin{itemize}
        \item \textit{Static examples (closed-book):} Few-shot examples are fixed and manually selected based on prior experimentation to maximize success. These remain the same across evaluations.
        
        \item \textit{Dynamic retrieval (open-book):} For each bug report, two similar examples are retrieved using BM25 based on textual similarity. These are drawn from the same dataset and tailored to the query, testing whether SUEs can leverage closely related context.
    \end{itemize}
    \
    Open-book evaluations were conducted only on the mutated versions of the benchmark: those using meaningful masking, rephrasing, and the combination of both. These settings were chosen to test whether SUEs can benefit from additional context when the original method names or bug report phrasing have been altered.
    \\
    To evaluate how different parts of a bug report influence test case generation, we manually annotated reports in the Defects4J dataset. Each report was labeled with the presence or absence of specific components such as test code, stack traces, method names, and natural language descriptions. The full list of components and annotation guidelines is provided in Table~\ref{table:component_labels}.
    \\
    \textbf{Evaluation Scheme:} 
    We track the percentage of bugs with BRTs in each setting to measure the effect of open-book vs. closed-book examples. For component-level analysis, we train 10 Random Forest classifiers with an 80–20 train-test split to evaluate the predictive power of individual bug report components, and aggregate the results. We apply the Scott-Knott test to identify statistically significant differences in component importance across models and mutation types.

\begin{table}[H]
    \centering
    \resizebox{\columnwidth}{!}{%
    \renewcommand{\arraystretch}{1.5} % Adjust row height slightly
    {\footnotesize % Reduce font size for the table
    \begin{tabular}{p{0.32\columnwidth} p{0.42\columnwidth} c} % Adjusted column widths
        \textbf{\textit{Component}}                         & \textbf{\textit{Label Guideline}}                                                                                 & \textbf{\textit{Presence (\%)}}\\ \hline
        \textbf{\textit{Natural Language Bug Description}}  & Explanation of where the system breaks is given                                                                   & 92.75\% \\ \hline
        \textbf{\textit{Buggy Method API Name}}             & The name of any of the patched methods is included in the bug report                                              & 63.75\% \\ \hline
        \textbf{\textit{Natural Language Bug Fix}}          & Without providing actual code, the developer mentioned what a possible fix could be to resolve the issue          & 3.93\% \\ \hline
        \textbf{\textit{Bug Fix Code}}                      & A diff or code block is provided with changes needed to be made or a complete method implementation               & 5.15\% \\ \hline
        \textbf{\textit{Developer’s Intent}}                & Details/Steps/Code given that led the developer to experience the bug                                             & 6.51\% \\ \hline
        \textbf{\textit{Test Code}}                         & Partial or complete Java test code (according to the developer) given                                             & 46.92\% \\ \hline
        \textbf{\textit{Expected/Actual Output}}            & Provided the expected output or the actual output                                                                 & 26.04\% \\ \hline
        \textbf{\textit{Stack Trace}}                       & The stack trace is attached in the bug report                                                                     & 10.93\% \\
    \end{tabular}
    }
    }
    \caption{Bug Report Components and Label Guideline}
    \label{table:component_labels}
\end{table}

\subsection{Evaluate}
    \textbf{Objective:}
    The goal of evaluation at this layer is to assess whether the SUEs can identify the most effective and non-redundant test cases from a set of generated candidates.
    \\
    \textbf{Approach:} \\
    Beyond generating test cases, effective test automation also requires the ability to evaluate and prioritize outputs. In this layer, we investigate whether the SUEs can distinguish between redundant, weak, or ineffective test cases and those that are more meaningful and aligned with the bug report. Unlike previous layers, the focus here shifts to selection: identifying the most promising test case from a set of generated candidates.
    \\
    To simulate realistic conditions, we utilize execution results on the buggy version of the codebase only. Our selection strategy builds on the LIBRO pipeline:
    \
    \begin{itemize}
        \item \textit{Cluster by output similarity:} Test cases are grouped based on the similarity of their error messages. Clusters that closely match the bug report description are prioritized.
        
        \item \textit{Filter by cluster size:} Small clusters are removed under the assumption that agreement across multiple generations signals higher reliability.
        
        \item \textit{Group by syntax:} Within each remaining cluster, we group test cases by syntactic equivalence to eliminate minor variations.
        
        \item \textit{Select diverse representatives:} From each group, we retain a syntactically diverse set of test cases to present to the SUE.
    \end{itemize}
    \
    This filtered set is then presented to the SUE, which is tasked with selecting the most effective test case. This final decision is critical, as some test cases may expose the bug only in the buggy version but fail to validate the fix, while others may offer broader coverage and better assertive structure. By evaluating the model’s ability to rank and choose among diverse candidates, we assess its internal reasoning capabilities and judgment when execution feedback is presented but limited.
    \\
    \textbf{Evaluation Scheme:}
    We adopt evaluation metrics introduced by the LIBRO framework to assess the quality of test case selection from a set of candidates:
    \
    \begin{itemize}
        \item \textit{Suggestion precision:} The proportion of suggested test cases that are valid BRTs.
        
        \item \textit{Suggestion recall:} The proportion of reproducible bugs among all BRTs.
        
        \item \textit{Accuracy@n (acc@n):} The proportion of bugs for which at least one true BRT appears in the top-\textit{n} ranked suggestions.
        
        \item \textit{Wasted effort (wef@n):} For bugs with at least one BRT in the top-\textit{n} ranked suggestions, this measures the number of non-BRTs encountered before the first valid one.
    \end{itemize}
    \
    These metrics evaluate the SUE’s ability to reason about and select effective test cases under realistic execution constraints.

\subsection{Create}
    \textbf{Objective:}
    The goal of evaluation at this layer is to assess whether SUEs can recognize ambiguity in bug reports and proactively seek clarification from the bug reporter before attempting bug reproduction.
    \\
    \textbf{Approach:}
    In practice, many bug reports lack clarity or include incomplete information. Addressing these cases requires models not just to interpret and reason, but to identify missing context and request further details. This layer targets the ability to detect ambiguity and generate meaningful clarification questions. Unlike the previous layers, which focused on generation and selection, this layer shifts the emphasis toward interactive reasoning.  
    \\
    Existing SUEs operate in a single-pass mode without mechanisms for querying or verifying assumptions. As such, we leave the evaluation of this capability for future work and consider it a forward-looking direction in automated test generation.
    \\
    \textbf{Evaluation Scheme:}
    We propose the following metrics to evaluate the effectiveness of ambiguity detection and clarification generation in future SUEs:
    \
    \begin{itemize}
        \item \textit{Ambiguity detection rate:} Percentage of unclear bug reports for which the model flags uncertainty or identifies missing information.
                
        \item \textit{Reproduction improvement rate:} The improvement in BRT success rate when clarification is provided and incorporated, compared to attempts without clarification.
    \end{itemize}

\section{Case Study}
\label{sec:casestudy}
In the previous section, we introduced our framework for systematically assessing the capabilities of LLM-powered Test Case Generation solutions. In this section, we present our case study on evaluating LIBRO, a SOTA test case generation framework. LIBRO utilizes a few-shot prompting strategy to generate test cases and inject them carefully inside the appropriate location in the repository. We selected two LLMs for evaluation based on their popularity and availability: (i) StarCoder-15.5B, an open-source model with publicly available training data and pipeline, specialized for code generation; (ii) a SOTA closed-source model trained by OpenAI. For simplicity, we will refer to these models as StarCoder and GPT throughout the rest of the paper. The detailed evaluation results for each layer will be discussed in the remaining part of this section. To save space, we abbreviate the results for the above four possible setups (two Benchmarks × two LLMs) evaluated at different layers as: {\textit{Benchmark}}\(\substack{\textit{\scriptsize Layer} \\ \textit{\scriptsize LLM}}\) in the remaining part of this section.

\subsection{Remember}
    Table~\ref{table:remember_results} presents the \%BRT results for the Remember layer across the four evaluation setups. \textit{Defects4J}\(\substack{\textit{\scriptsize Remember} \\ \textit{\scriptsize StarCoder}}\) achieves 24.1\%, closely aligning with the 23.6\% reported in the original LIBRO study. \textit{Defects4J}\(\substack{\textit{\scriptsize Remember} \\ \textit{\scriptsize GPT}}\) attains 36.5\%, outperforming the earlier Codex result of 31.5\%. On the GHRB benchmark, \textit{GHRB}\(\substack{\textit{\scriptsize Remember} \\ \textit{\scriptsize GPT}}\) maintains parity with Codex at 32.3\%, while \textit{GHRB}\(\substack{\textit{\scriptsize Remember} \\ \textit{\scriptsize StarCoder}}\) drops to 16.1\%, showing a considerable decrease from the originally reported 29.0\%. These findings suggest that GPT replicates or improves upon prior results across both benchmarks, while StarCoder demonstrates more consistent performance on Defects4J than on GHRB.
    
    To better understand these trends, we consider potential training data contamination. Both benchmarks include real-world bugs from popular Java projects that are likely part of LLM training corpora. StarCoder was trained on \textit{The Stack} dataset with a cut-off date of June 2022~\cite{Kocetkov_2022_TheStack}, which includes repositories from the Defects4J benchmark, as confirmed through the \textit{Am I In The Stack?}\footnote{https://huggingface.co/spaces/bigcode/in-the-stack} platform. GPT-4o was trained on data available up to October 2023~\cite{OpenAI_2024_GPT4o}. In comparison, the latest bug fix in Defects4J was committed in July 2019, while GHRB covers bugs from July 2021 to July 2022. These timelines indicate that both StarCoder and GPT may have had partial exposure to these benchmarks during training, potentially influencing their ability to reproduce bug-revealing test cases under unchanged conditions.

    \begin{summarybox}
        \textbf{Summary:} \textit{Defects4J}\(\substack{\textit{\scriptsize Remember} \\ \textit{\scriptsize StarCoder}}\) and \textit{GHRB}\(\substack{\textit{\scriptsize Remember} \\ \textit{\scriptsize GPT}}\) match previously reported results, while \textit{Defects4J}\(\substack{\textit{\scriptsize Remember} \\ \textit{\scriptsize GPT}}\) exceeds its predecessor, and \textit{GHRB}\(\substack{\textit{\scriptsize Remember} \\ \textit{\scriptsize StarCoder}}\) underperforms. These findings highlight that even within familiar benchmarks, performance can fluctuate based on model type and benchmark-specific factors.
    \end{summarybox}

    \begin{table}[h]
    \centering
    \caption{\%BRT results for the \textsc{Remember} layer.}
    \label{table:remember_results}
    \begin{tabular}{@{}llcc@{}}
    \toprule
    \textbf{Benchmark} & \textbf{Model} & \textbf{Ours} & \textbf{Original} \\
    \midrule
    \multirow{2}{*}{Defects4J}
        & StarCoder       & 24.1\% & 23.6\% \\
        & Codex/GPT-4o    & 36.5\% & 31.5\% \\
    \midrule
    \multirow{2}{*}{GHRB}
        & StarCoder       & 16.1\% & 29.0\% \\
        & Codex/GPT-4o    & 32.3\% & 32.3\% \\
    \bottomrule
    \end{tabular}
    \end{table}

\subsection{Understand}
    To evaluate whether SUEs can generalize beyond the exact phrasing recorded in the original issue trackers, we assessed performance on bug reports that were rephrased or translated into Romanized Urdu. The \%BRT results are summarized in Table~\ref{table:understand_results}. On the \textit{Defects4J} benchmark, \textit{Defects4J}\(\substack{\textit{\scriptsize Understand} \\ \textit{\scriptsize StarCoder}}\) yields a \%BRT of 21.1\% on rephrased bug reports and 20.6\% on translated ones, reflecting a modest drop relative to the original 24.1\%. In contrast, \textit{Defects4J}\(\substack{\textit{\scriptsize Understand} \\ \textit{\scriptsize GPT}}\) records \%BRTs of 35.5\% and 37.2\%, closely aligning with its original performance of 36.5\%. On the \textit{GHRB} benchmark, both models maintain similar performance across linguistic variations, with \textit{GHRB}\(\substack{\textit{\scriptsize Understand} \\ \textit{\scriptsize StarCoder}}\) achieving 16.1\% and 22.6\%, and \textit{GHRB}\(\substack{\textit{\scriptsize Understand} \\ \textit{\scriptsize GPT}}\) achieving 32.3\% and 41.9\%. These results indicate that GPT is more resilient to linguistic variation, whereas StarCoder shows greater sensitivity to changes in input phrasing.
    
    The \textit{Baseline Consistency Rate} results reinforce the models’ ability to retain previously generated test cases across linguistic variations. On \textit{Defects4J}, StarCoder records consistency rates of 70.9\% and 68.4\% for rephrased and translated inputs, respectively, while GPT achieves higher rates of 83.8\% and 87.8\%. On \textit{GHRB}, StarCoder reaches 60.0\% and 80.0\%, whereas GPT maintains 90.0\% and 100\% over both conditions. While these rates indicate significant overlap with the original \%BRT outcomes, the overall \%BRT does not decline substantially which suggests that some bugs are newly reproduced in the modified settings. This indicates that linguistic variation can occasionally enable successful test generation for cases that were previously unsolved. Rather than reflecting a strict loss in performance, these results highlight the models’ capacity to generalize and adapt to alternative phrasings.

    We applied McNemar’s test comparing the baseline setup to each modified condition. Most comparisons yielded $p$-values above the standard significance threshold ($p \geq 0.05$), indicating no statistically significant shift in reproduction outcomes. However, for \textit{Defects4J}\(\substack{\textit{\scriptsize Understand} \\ \textit{\scriptsize StarCoder}}\), both the rephrased ($p = 0.008$) and translated ($p = 0.001$) conditions exhibit statistically significant differences. These results suggest that StarCoder’s behavior is notably influenced by changes in linguistic presentation, while GPT remains more robust under such variations.
    
    \begin{summarybox}
        \textbf{Summary:} GPT maintains robust performance under both rephrased and translated bug report conditions, with minimal changes in \%BRT and high Baseline Consistency Rates. StarCoder, in contrast, exhibits more sensitivity to linguistic variations, with statistically significant drops in reproduction outcomes on \textit{Defects4J}. Despite these shifts, both models demonstrate the ability to generalize across alternative phrasings, as evidenced by stable overall performance and newly reproduced bugs in modified settings. These findings highlight that modern SUEs are not only capable of retaining learned representations but can also adapt flexibly to variations in bug report formulations.
    \end{summarybox}

    \begin{table}[h]
    \centering
    \caption{Performance of SUEs under linguistic mutations in the \textsc{Understand} layer.}
    \label{table:understand_results}
    \begin{tabular}{llccc}
    \toprule
    \textbf{Benchmark} & \textbf{Model} & \textbf{Mutation} & \textbf{\%BRT} & \textbf{BCR} \\
    \midrule
    \multirow{4}{*}{Defects4J} 
        & \multirow{2}{*}{StarCoder} 
            & Rephrased & 21.1\% & 70.9\% \\
        &  & Translated & 20.6\% & 68.4\% \\
        \cmidrule{2-5}
        & \multirow{2}{*}{GPT-4o} 
            & Rephrased & 35.5\% & 83.8\% \\
        &  & Translated & 37.2\% & 87.8\% \\
    \midrule
    \multirow{4}{*}{GHRB} 
        & \multirow{2}{*}{StarCoder} 
            & Rephrased & 16.1\% & 60.0\% \\
        &  & Translated & 22.6\% & 80.0\% \\
        \cmidrule{2-5}
        & \multirow{2}{*}{GPT-4o} 
            & Rephrased & 32.3\% & 90.0\% \\
        &  & Translated & 41.9\% & 100.0\% \\
    \bottomrule
    \end{tabular}%
    \end{table}

\subsection{Apply}
    To evaluate whether SUEs can apply learned behavior when identifier information is altered, we test each model under four mutation strategies: \textit{Hash Mask}, \textit{Meaningful Mask}, \textit{Meaningful Mask + Rephrased}, and \textit{Meaningful Mask + Urdu}. The \%BRT results under these conditions are summarized in Table~\ref{table:apply_results}. On the \textit{Defects4J} benchmark, \textit{Defects4J}\(\substack{\textit{\scriptsize Apply} \\ \textit{\scriptsize StarCoder}}\) drops sharply from a baseline of 24.1\% to 6.9\% with Hash Mask and 10.7\% with Meaningful Mask. The rephrased and Urdu variants yield slightly higher rates (11.5\% and 12.7\%, respectively) but remain well below baseline. \textit{Defects4J}\(\substack{\textit{\scriptsize Apply} \\ \textit{\scriptsize GPT}}\) also declines from 36.5\% to 22.4\% (Hash Mask) and 21.7\% (Meaningful Mask), with rephrased and Urdu variants dropping further to 19.8\% and 19.7\%. A similar trend holds on the \textit{GHRB} benchmark, where both models perform noticeably worse under all masking conditions, with the best results observed under meaningful masking.
    
    We measure how much of the baseline reproduction performance is preserved using the Baseline Consistency Rate. For StarCoder on \textit{Defects4J}, this value falls to 24.0\% with Hash Mask and remains around 37–40\% for the meaningful and its variants. GPT maintains higher consistency, between 47–52\% across all mutations. These figures highlight the models’ strong dependence on identifier-level cues. Although a few bugs not reproduced in the baseline are resolved under masking, these gains are small relative to the overall performance loss. This suggests that while the models may occasionally generalize to new reasoning paths, identifier mutations primarily disrupt rather than enhance test generation.
    
    We applied McNemar’s test to assess whether identifier mutations lead to statistically significant changes in bug reproduction outcomes. On \textit{Defects4J}, both StarCoder and GPT exhibit significant differences under Hash Mask and Meaningful Mask conditions ($p$-values $<$ 0.05). On \textit{GHRB}, similar results hold for GPT, while StarCoder does not show significant changes, which is likely due to the smaller number of reproducible cases. These findings confirm that altering identifier information affects model behavior significantly.
    
    \begin{summarybox}
        \textbf{Summary:} Identifier mutations cause substantial performance declines for both SUEs, with StarCoder more severely impacted than GPT. Although a few new bugs are occasionally reproduced under masked settings, these gains are limited and do not compensate for the broader loss in consistency. Baseline Consistency Rates and McNemar test results indicate that both models remain strongly reliant on familiar naming patterns. This limitation highlights a core challenge in achieving generalization across codebases with unfamiliar or altered identifier conventions.
    \end{summarybox}

    \begin{table}[h]
    \centering
    \caption{Performance of SUEs under identifier mutations in the \textsc{Apply} layer.}
    \label{table:apply_results}
    \begin{tabular}{llccc}
    \toprule
    \textbf{Benchmark} & \textbf{Model} & \textbf{Mutation} & \textbf{\%BRT} & \textbf{BCR} \\
    \midrule
    \multirow{8}{*}{Defects4J} 
        & \multirow{4}{*}{StarCoder} 
            & Hash Mask & 6.9\% & 24.0\% \\
        &  & Meaningful Mask & 10.7\% & 38.3\% \\
        &  & Meaningful Mask + Rephrased & 11.5\% & 37.2\% \\
        &  & Meaningful Mask + Urdu & 12.7\% & 40.3\% \\
        \cmidrule{2-5}
        & \multirow{4}{*}{GPT-4o} 
            & Hash Mask & 22.4\% & 52.7\% \\
        &  & Meaningful Mask & 21.7\% & 51.0\% \\
        &  & Meaningful Mask + Rephrased & 19.8\% & 47.3\% \\
        &  & Meaningful Mask + Urdu & 19.7\% & 47.0\% \\
    \midrule
    \multirow{8}{*}{GHRB} 
        & \multirow{4}{*}{StarCoder} 
            & Hash Mask & 3.2\% & 20.0\% \\
        &  & Meaningful Mask & 9.7\% & 40.0\% \\
        &  & Meaningful Mask + Rephrased & 6.5\% & 40.0\% \\
        &  & Meaningful Mask + Urdu & 3.2\% & 20.0\% \\
        \cmidrule{2-5}
        & \multirow{4}{*}{GPT-4o} 
            & Hash Mask & 9.7\% & 30.0\% \\
        &  & Meaningful Mask & 12.9\% & 40.0\% \\
        &  & Meaningful Mask + Rephrased & 12.9\% & 30.0\% \\
        &  & Meaningful Mask + Urdu & 16.1\% & 40.0\% \\
    \bottomrule
    \end{tabular}%
    \end{table}

\subsection{Analyze}
\begin{figure}[htbp]
        \centering
        \includegraphics[width=\columnwidth]{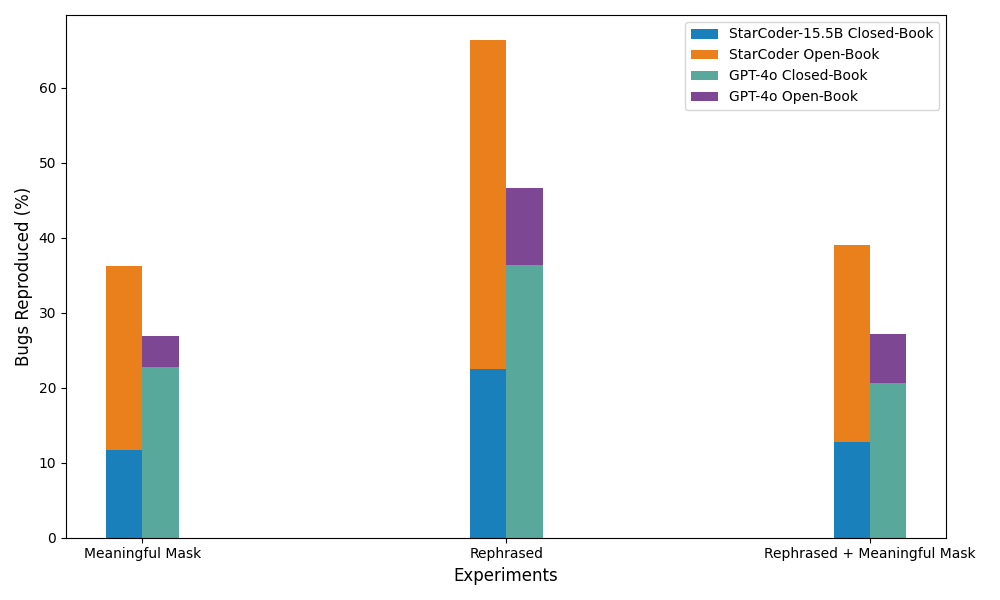}
        \caption{Open-vs-Closed Book Comparison for Defects4J Benchmark}
        \label{fig:openbook_vs_closedbook_d4j}
    \end{figure}

\begin{figure}[htbp]
        \centering
        \includegraphics[width=\columnwidth]{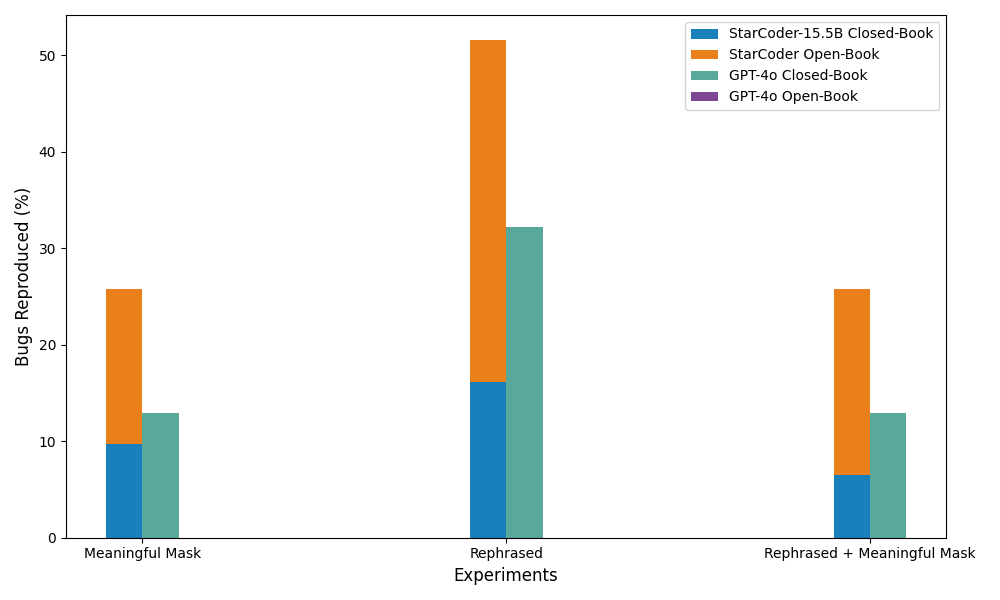}
        \caption{Open-vs-Closed Book Comparison for GHRB Benchmark}
        \label{fig:openbook_vs_closedbook_ghrb}
    \end{figure}

    To evaluate how SUEs analyze and apply contextual information, we compare performance in closed-book and open-book settings. The \%BRT results for open-book and closed-book configurations are summarized in Table~\ref{table:open_closed_results} and visualized in Figure~\ref{fig:openbook_vs_closedbook_d4j} and Figure~\ref{fig:openbook_vs_closedbook_ghrb}. On the \textit{Defects4J} benchmark, open-book prompting yields improvements ranging from 209.9\% to 229.0\% for StarCoder and from 18.2\% to 29.2\% for GPT across different mutation types. On \textit{GHRB}, the gains are even more significant for StarCoder, ranging from 166.7\% to 300.0\%, while GPT shows no measurable improvement. These findings highlight the ability of LLMs to analyze near-identical retrieved examples and apply in-context learning to mutated bug reports. Compared to closed-book setup, the open-book setting provides more targeted and adaptable context, enabling better generalization when familiar identifiers or phrasings are missing.

    To further assess how SUEs interpret bug report content itself, we conducted a component-level analysis using annotated bug reports from the \textit{Defects4J} dataset. We trained 10 Random Forest classifiers to predict reproduction success based on the presence of various structured and unstructured components. We used the Scott-Knott test on their aggregated results to rank their relative importance across different mutation conditions (Figure~\ref{fig:scott_knott_main}). Full results for all mutation conditions are provided in Appendix~\ref{appendix:scott_knott}.

    The results show that LLMs rely most heavily on structured technical components such as test code, method names, and stack traces. These trends remain consistent across all mutation conditions, suggesting that when identifier-based cues are obscured, SUEs shift their attention toward more concrete, executable signals. In contrast, natural language descriptions and high-level fix explanations contribute relatively little to reproduction success. This indicates that current SUEs are less capable of translating narrative information into actionable behavior, underscoring a key limitation in their ability to reason from unstructured, descriptive content.
    
    \begin{summarybox}
        \textbf{Summary:} Open-book prompting substantially improves test case generation by enabling LLMs to apply knowledge from retrieved, contextually similar examples. Component-level analysis further reveals that models extract the most useful signals from structured technical content, with test code, buggy method name and stack traces being of highest importance. These findings suggest that successful test generation hinges on the model’s ability to analyze and adapt both external examples and internal report structures.
    \end{summarybox}

    \begin{table}[ht]
    \centering
    \caption{\%BRT in closed-book and open-book settings across models, benchmarks, and mutation types.}
    \label{table:open_closed_results}
    \begin{tabular}{llccc}
    \toprule
    \textbf{Benchmark} & \textbf{Model} & \textbf{Mutation} & \textbf{\%BRT (Closed)} & \textbf{\%BRT (Open)} \\
    \midrule
    \multirow{6}{*}{Defects4J} 
    & \multirow{3}{*}{StarCoder} 
      & Meaningful Mask           & 10.7\% & 34.9\% \\
    & & Rephrased                  & 22.5\% & 65.3\% \\
    & & Meaningful + Rephrased     & 11.5\% & 37.7\% \\[0.5ex]
    \cmidrule{2-5}
    & \multirow{3}{*}{GPT-4o} 
      & Meaningful Mask           & 21.7\% & 25.6\% \\
    & & Rephrased                  & 36.3\% & 45.3\% \\
    & & Meaningful + Rephrased     & 19.8\% & 25.6\% \\
    \midrule
    \multirow{6}{*}{GHRB} 
    & \multirow{3}{*}{StarCoder} 
      & Meaningful Mask           & 9.7\%  & 25.8\% \\
    & & Rephrased                  & 16.1\% & 51.6\% \\
    & & Meaningful + Rephrased     & 6.5\%  & 25.8\% \\[0.5ex]
    \cmidrule{2-5}
    & \multirow{3}{*}{GPT-4o} 
      & Meaningful Mask           & 12.9\% & 12.9\% \\
    & & Rephrased                  & 32.3\% & 32.3\% \\
    & & Meaningful + Rephrased     & 12.9\% & 12.9\% \\
    \bottomrule
    \end{tabular}
    \end{table}

    \begin{figure}[H]
      \centering
      \includegraphics[width=\columnwidth]{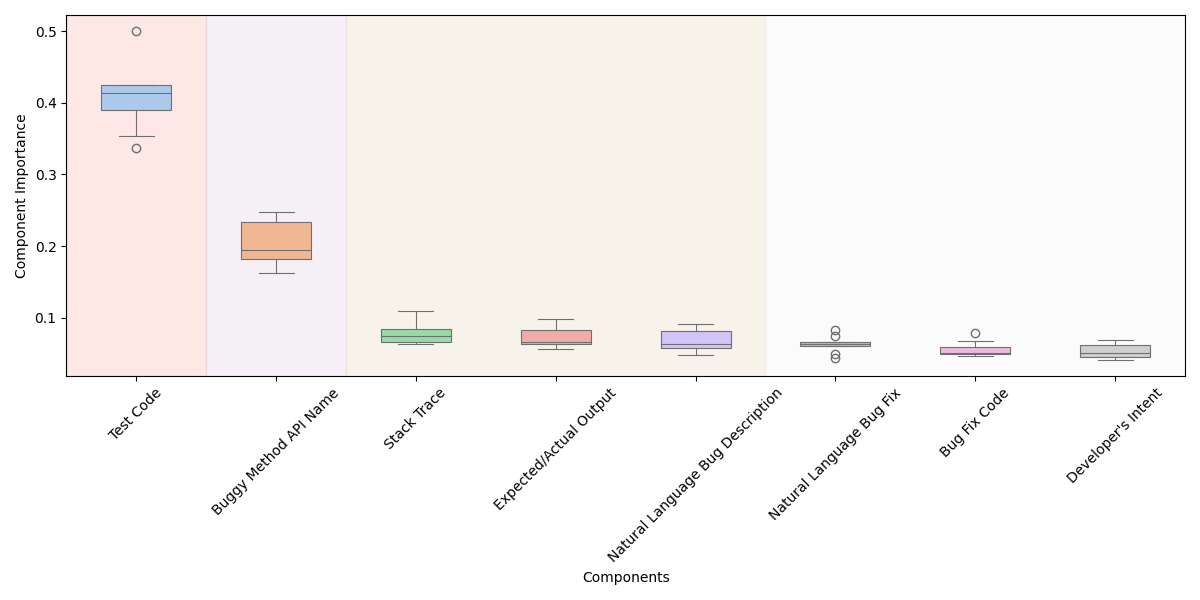}
      \caption{Scott-Knott ranks of bug report components on \textit{Defects4J}.}
      \label{fig:scott_knott_main}
    \end{figure}

\begin{figure*}[!h]
  \centering
  \includegraphics[width=\textwidth]{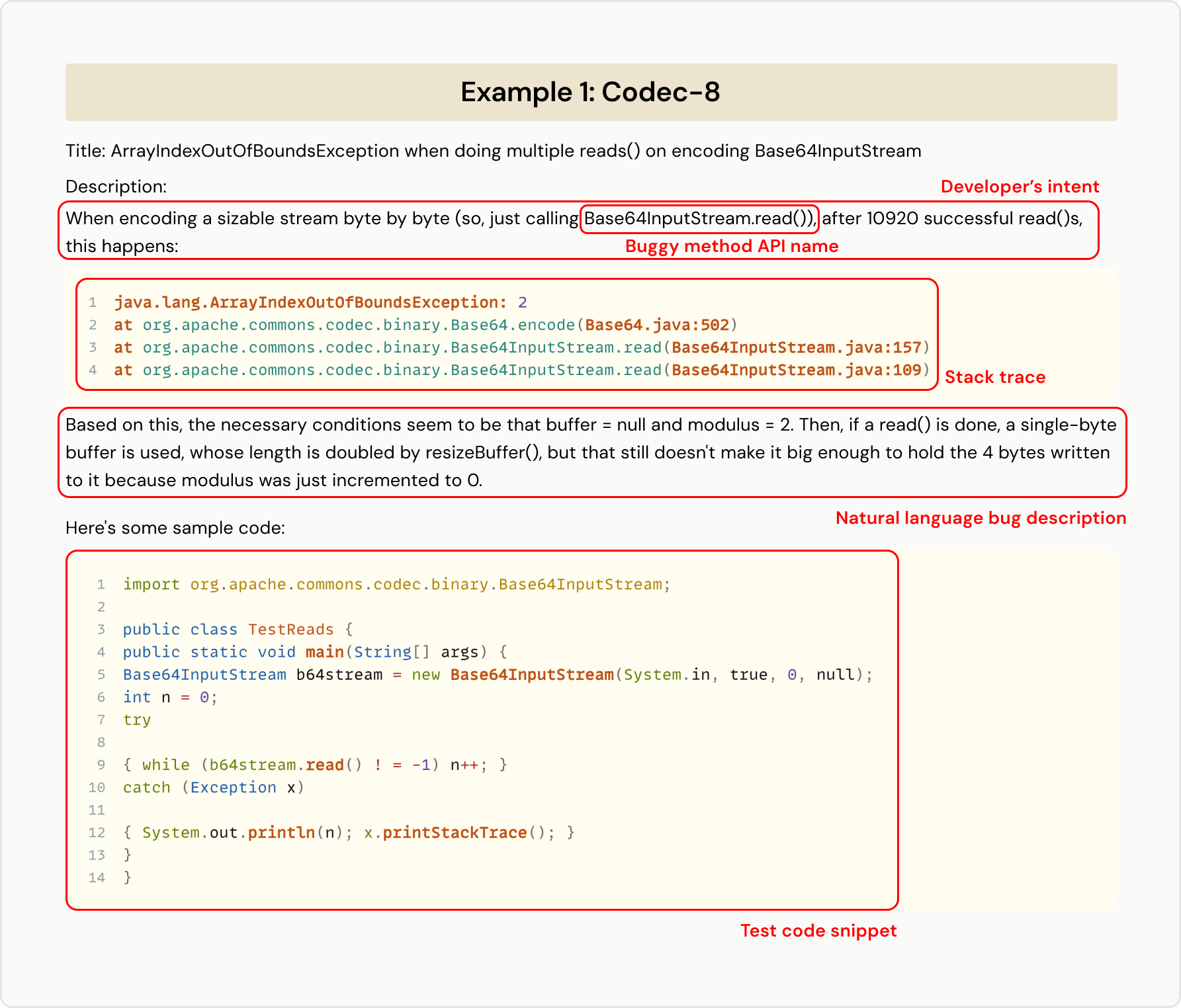}
  \label{fig:annotation_example_1}
\end{figure*}

\newpage
\begin{figure*}[!t]
  \centering
  \includegraphics[width=\textwidth]{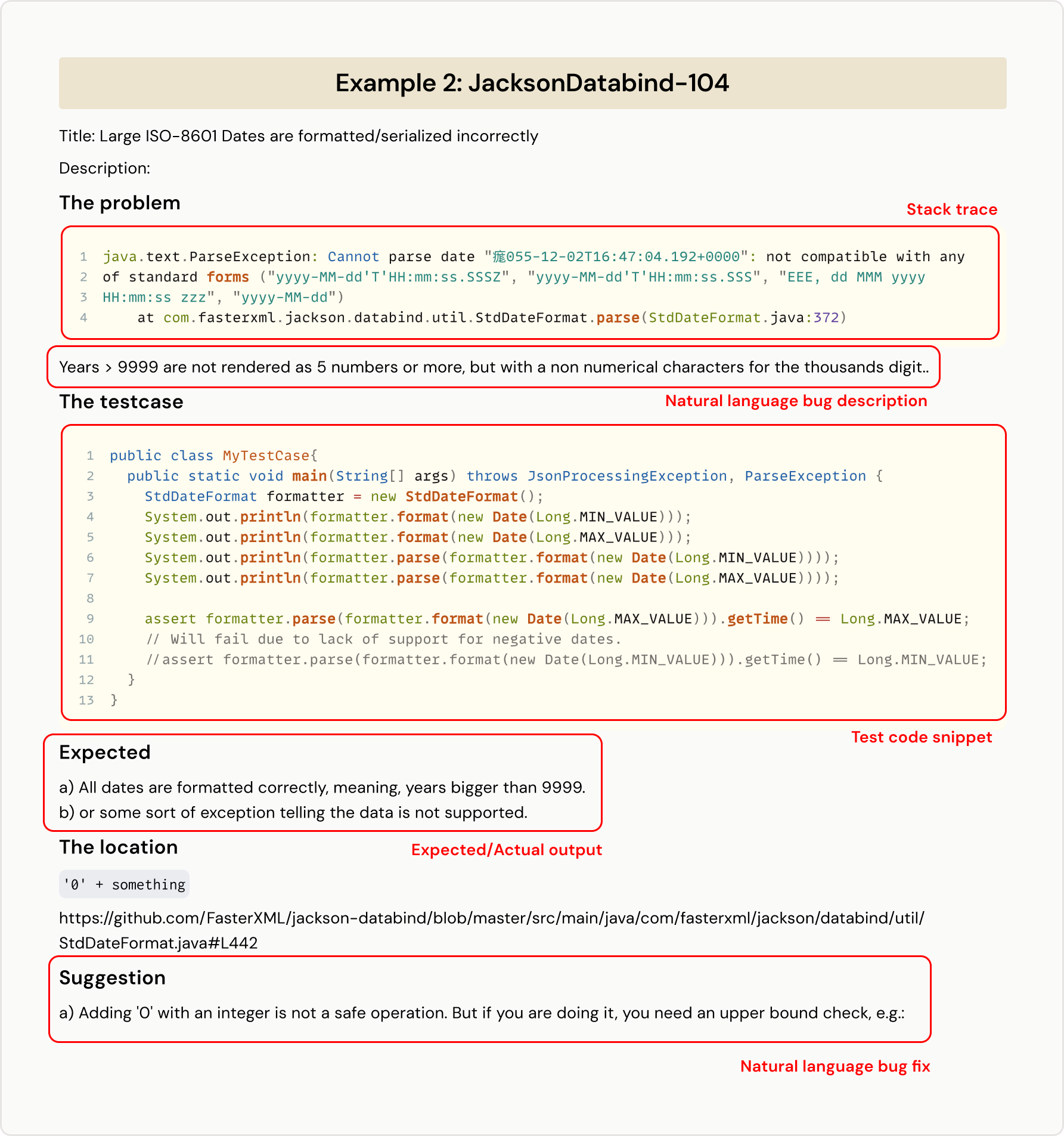}
  \label{fig:annotation_example_2a}
\end{figure*}

\newpage
\begin{figure*}[!t]
  \centering
  \includegraphics[width=\textwidth]{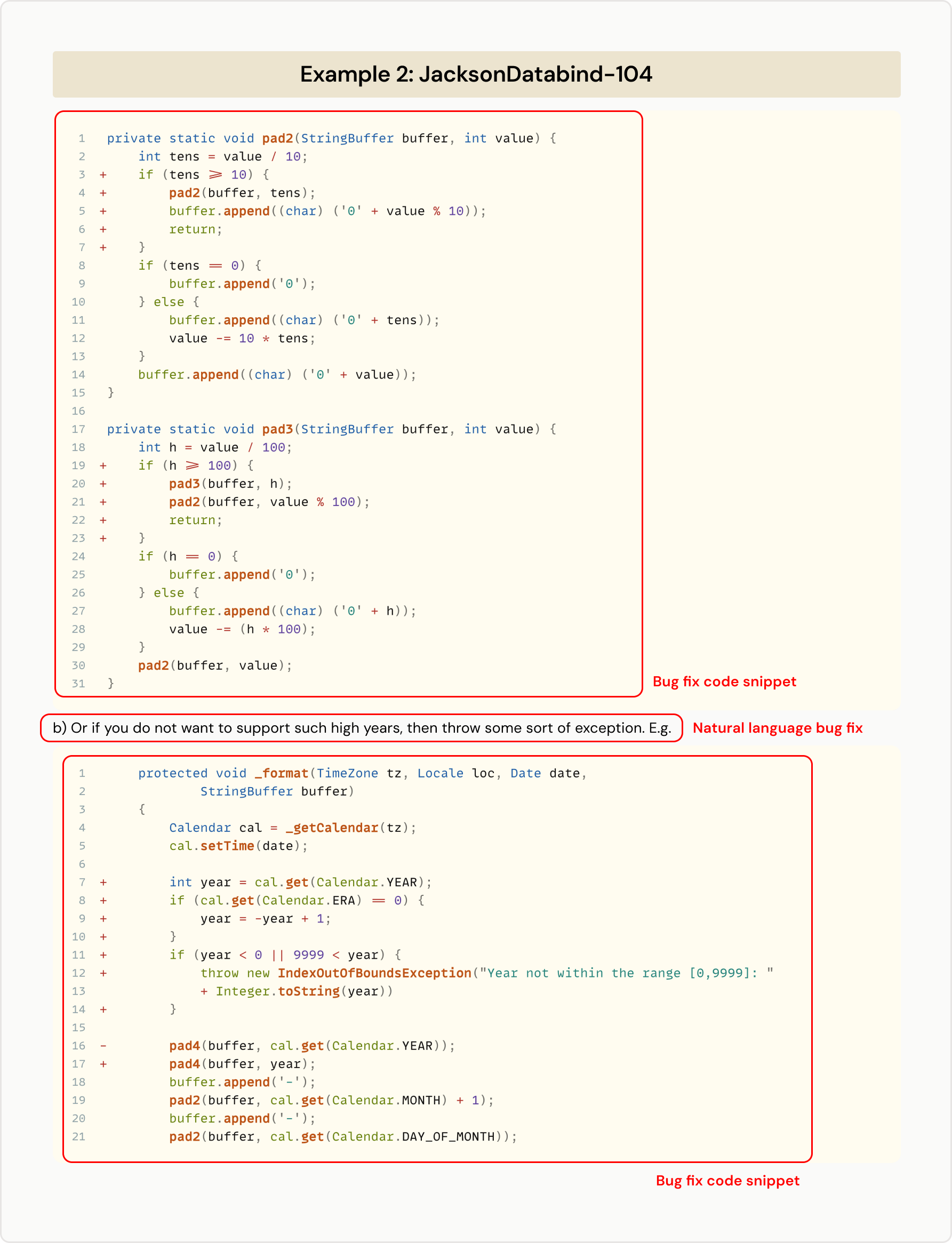}
  \label{fig:annotation_example_2b}
\end{figure*}

\subsection{Evaluate}

\section{Discussion}
\label{sec:discussion}
\subsection{Benchmark Quality}
    The reliability of benchmark datasets plays a critical role in assessing LLMs' generalization capabilities. Our analysis of the 812 bug reports in Defects4J revealed notable concerns regarding the quality of available bug descriptions. Specifically, 47 bug reports lacked relevant information, making it difficult for both human developers and LLMs to generate meaningful test cases. These bug reports usually contained attached files with a bug fix patch and update patch for test suite as well. Additionally, we identified 29 groups of bugs linked to the same bug report, covering 66 distinct bugs in total. While these bugs represent different issues, their association with a single report limits the benchmark’s ability to evaluate LLMs on unique, independently described faults. These inconsistencies highlight the need for more structured, informative, and diverse benchmarks to ensure robust model evaluation.

\subsection{Comparison with Developer Survey done by Zimmermann}
    \subsubsection{Comparison with Developer Preferences}
        In addition to analyzing which bug report components are most useful to LLMs, we compare these findings with those from a prior study by Bettenburg et al.~\cite{Bettenburg_2010_WmaGBR}, which surveyed developers to understand which bug report elements they consider most important for successful bug reproduction. This survey collected importance rankings across multiple components, including stack traces, test cases, code examples, and natural language descriptions such as steps to reproduce and observed behavior.

        Our findings show a clear overlap between components valued by developers and those that are most predictive for LLMs. Structured and technical elements such as test code and stack traces are consistently ranked high in both analyses. Developers rated stack traces and test cases among the top three most useful components, and our model-level analysis confirms their importance in guiding successful test generation. These components provide executable and context-specific cues, which LLMs appear to use effectively when reasoning about reproduction steps.

        Despite several points of alignment, we observe key differences between LLM behavior and developer expectations. Developers rated descriptive elements such as steps to reproduce and observed behavior as highly important, ranking them first and fourth in the survey, respectively. However, our analysis shows that LLMs rely minimally on narrative or procedural descriptions. In our dataset, the closest corresponding component is “Developer’s Intent,” which captures high-level explanations of what the user was trying to do. These descriptions, while potentially useful, often lack structure and clarity. This likely limits their utility in automated reasoning.

        Interestingly, developers also report these same components as frequent sources of frustration. In the survey, 79\% cited errors in steps to reproduce as a major obstacle, while 48\% encountered incorrect descriptions of observed behavior. Moreover, 74\% of respondents reported missing or incomplete information, and over a third noted issues with unstructured or overly long text. These challenges may help explain why LLMs underutilize such components: even when present, they often lack the precision or consistency needed for effective test generation.
    
        This comparison highlights a fundamental gap between what developers need and what current LLMs are able to use effectively. Structured signals such as test code and stack traces remain the most actionable elements for both humans and models. In contrast, unstructured content, while essential in principle, is often underutilized due to its ambiguity or lack of clarity. Closing this gap may require improved LLM reasoning over loosely structured text, or interactive mechanisms that allow models to request missing or unclear information during the reproduction process.

% \subsection{Missing Components across Failing Bugs}
%     \irtaza{I dont think it was necessarily the missing components causing unsuccessful bug reproductions as explained in the previous subsection.}

\section{Conclusion}
\label{sec:conclusion}
This paper presents a structured evaluation framework for assessing LLM-based test generation, guided by Bloom’s Taxonomy. By applying systematic linguistic and contextual mutations to established benchmarks such as Defects4J and GHRB, we evaluate how well StarCoder and GPT-4o perform across the layers of \textit{Remember}, \textit{Understand}, \textit{Apply}, and \textit{Analyze}. Our findings reveal that while LLMs can recall previously seen tasks with high accuracy, they exhibit variability across repeated generations. GPT-4o demonstrates greater robustness than StarCoder when interpreting rephrased or translated bug reports, indicating stronger generalization under linguistic variation. However, both models experience a sharp decline when identifier information is obscured, highlighting a reliance on surface-level cues rather than deeper contextual understanding. When provided with relevant examples, both models benefit from alignment in technical details, supporting their ability to generalize under guidance. These results underscore the importance of evaluating LLMs beyond static benchmarks and offer a principled approach for identifying both capabilities and limitations in test case generation. Our framework provides a foundation for future assessments that aim to more faithfully reflect the challenges of real-world software engineering tasks.

\bibliographystyle{ACM-Reference-Format}
\bibliography{references}

\appendix

\section{Full Scott-Knott Results}
\label{appendix:scott_knott}
\begin{figure}[H]
  \centering
  \includegraphics[width=\columnwidth]{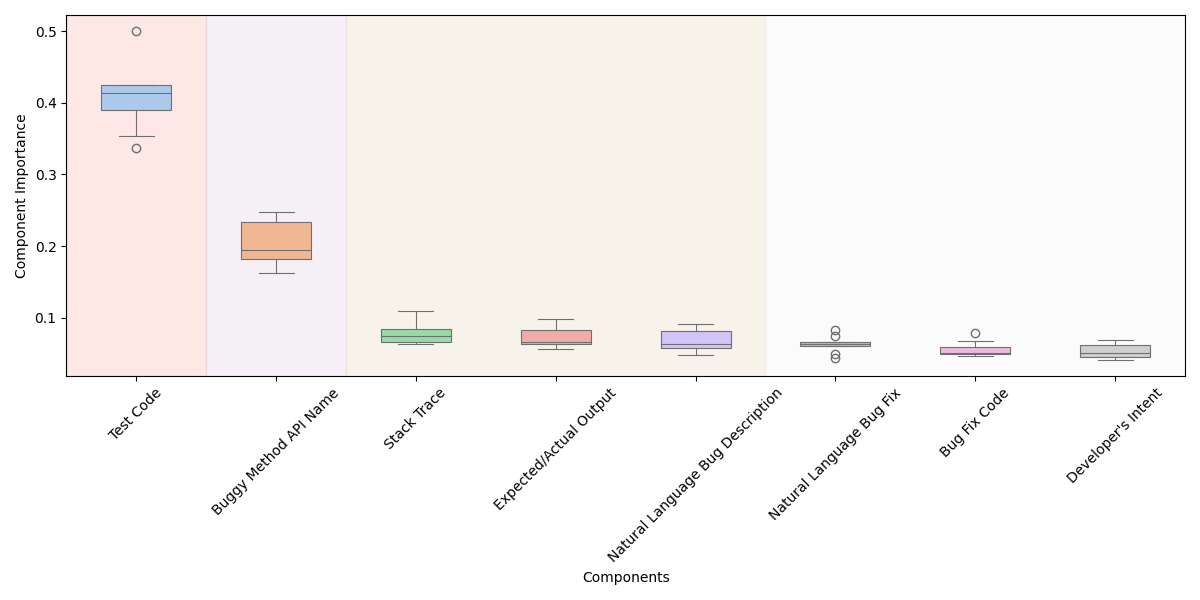}
  \caption{Scott-Knott ESD ranking of bug report components for the Baseline (LIBRO) using StarCoder on \textit{Defects4J}.}
\end{figure}

\begin{figure}[H]
  \centering
  \includegraphics[width=\columnwidth]{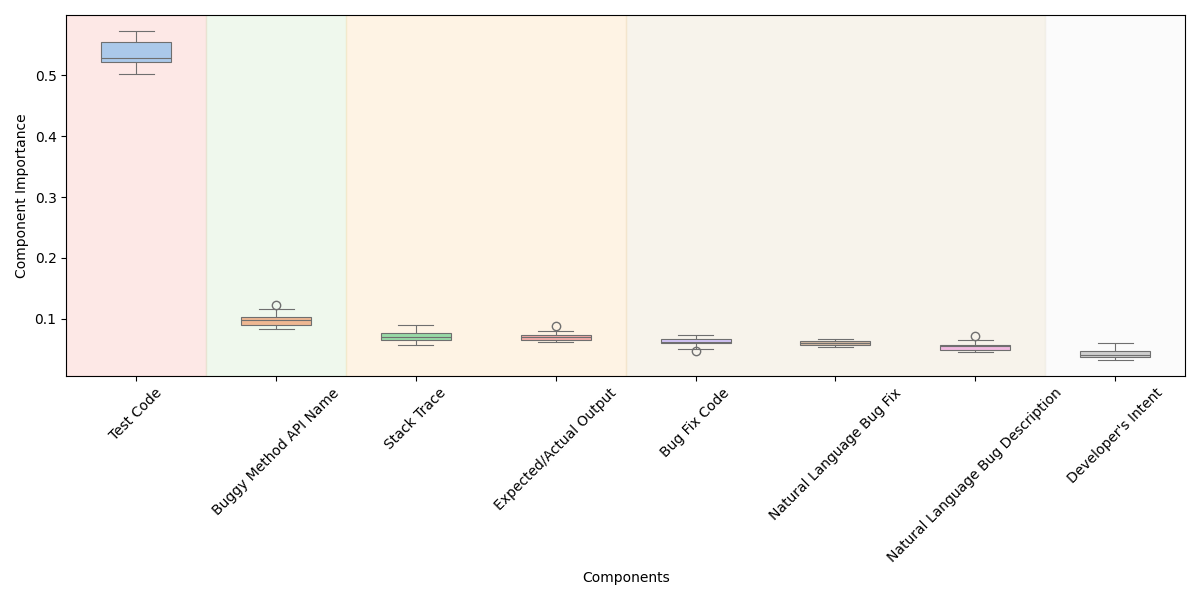}
  \caption{Scott-Knott ESD ranking of bug report components for the Rephrased + Meaningful Mask using StarCoder on \textit{Defects4J}.}
\end{figure}

\begin{figure}[H]
  \centering
  \includegraphics[width=\columnwidth]{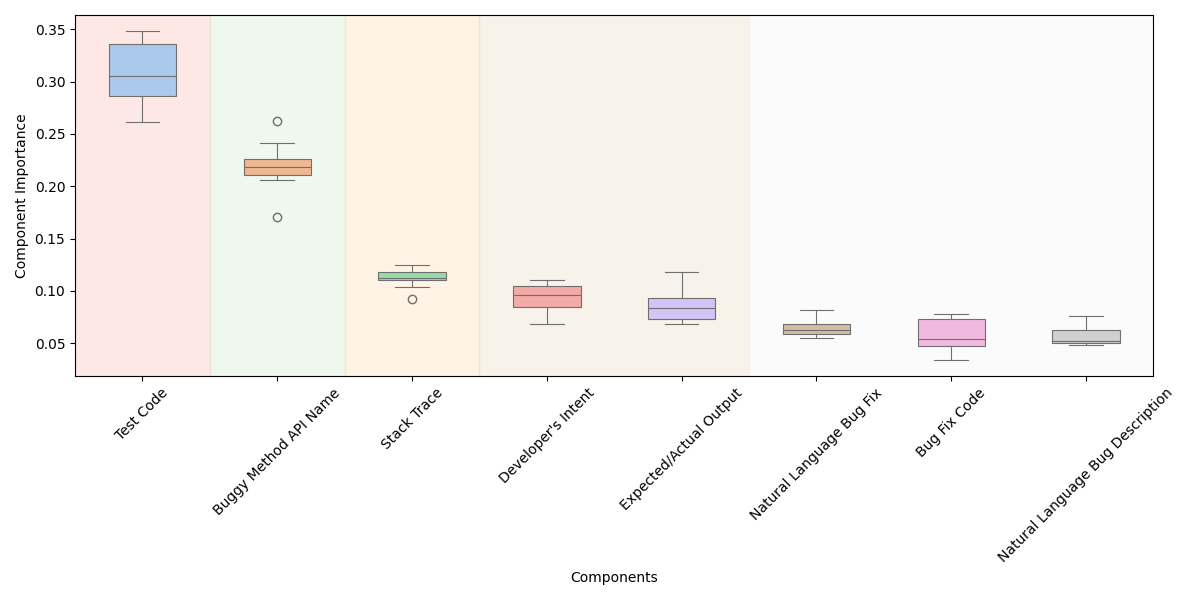}
  \caption{Scott-Knott ESD ranking of bug report components for the Baseline (LIBRO) using GPT-4o on \textit{Defects4J}.}
\end{figure}

\begin{figure}[H]
  \centering
  \includegraphics[width=\columnwidth]{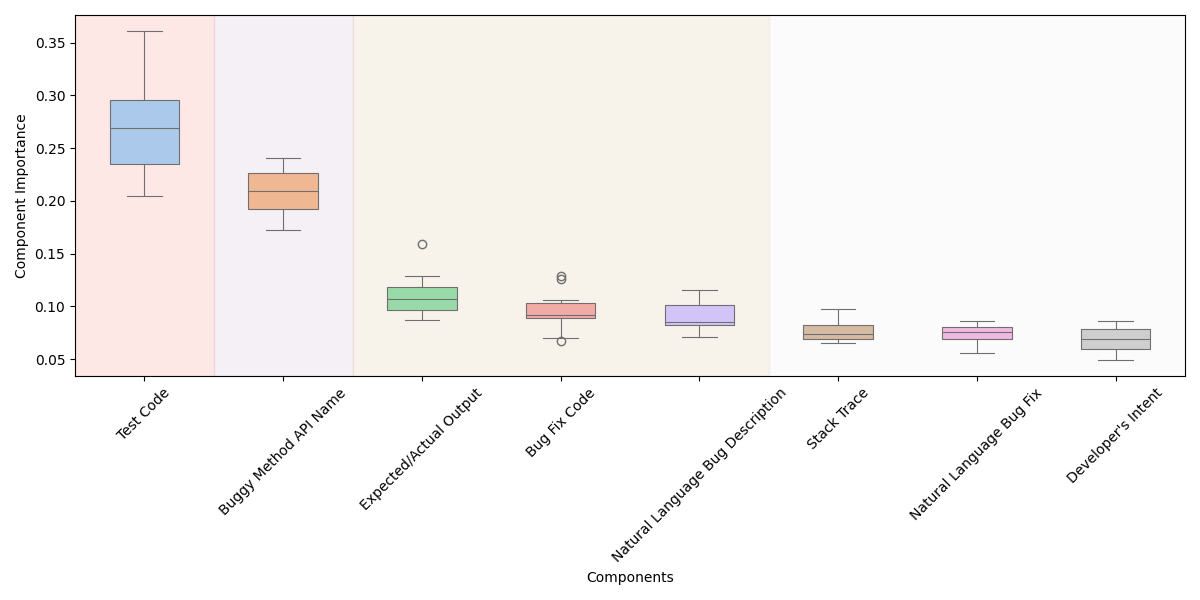}
  \caption{Scott-Knott ESD ranking of bug report components for the Rephrased + Meaningful Mask using GPT-4o on \textit{Defects4J}.}
\end{figure}

\end{document}